\documentclass[fleqn,twoside]{article}
\usepackage{espcrc2}

\usepackage{pslatex}
\usepackage{graphicx}
\usepackage[latin1]{inputenc}
\usepackage[T1]{fontenc}

\def\mhiggs{{m_H}}
\def\Fig#1{Fig.~\ref{#1}}

\def\Eq#1{Eq.~(\ref{#1})}
\def\Ref#1{Ref.~\cite{#1}}
\def\Refs#1{Refs.~\cite{#1}}
\def\ttbarjet{\ensuremath{t \bar t + \mbox{jet}}}
\def\nn{\nonumber}

\title{Top quark pair + jet production at next-to-leading order:\\ 
  NLO QCD corrections to $gg\to t\bar t g$}

\author{
  A.~Brandenburg\address{DESY-Theorie, 22603 Hamburg, Germany}%
  \thanks{Heisenberg Fellow of the
    Deutsche Forschungsgemeinschaft},
  S.~Dittmaier\address[MPI]{Max-Planck-Institut für Physik 
    (Werner-Heisenberg-Institut), Föhringer
    Ring 6, D-80805 Munich, Germany},
  P.~Uwer\address{Department of Physics, 
    TH Division, CERN, CH-1211 Geneva 23, 
    Switzerland
    },
  S.~Weinzierl\addressmark[MPI]\footnotemark[1]
  }
       
\begin{document} 
\setcounter{page}{0}
\thispagestyle{empty}
\begin{flushright}
  \large
  CERN-PH-TH/2004-122\\
  DESY 04-111\\
  MPP-2004-74
 \end{flushright}
\vspace*{4cm}

\begin{center}
  \Large
  {\bf Top quark pair + jet production at next-to-leading order:\\ 
  NLO QCD corrections to $gg\to t\bar t g$}
  \vspace*{2cm}
  
  A.~Brandenburg$^a$, 
  S.~Dittmaier$^b$, P.~Uwer$^c$, S.~Weinzierl$^b$\\[0.5cm]

  {\large $^a$DESY-Theorie, 22603 Hamburg, Germany\\[0.2cm]

  $^b$Max-Planck-Institut für Physik 
    (Werner-Heisenberg-Institut),\\ Föhringer
    Ring 6, D-80805 Munich, Germany\\[0.2cm]

  $^c$Department of Physics, 
    TH Division, CERN, CH-1211 Geneva 23, 
    Switzerland}\\[0.5cm]

  {\large \bf Abstract}\\[0.1cm]

  \parbox{16cm}{\large The reaction $pp/p\bar p \to \ttbarjet+X$ is an important background
  process for Higgs boson searches in the mass range below 200~GeV. Apart
  from that it is also an ideal laboratory for precision measurements in
  the top quark sector. Both applications require a solid theoretical
  prediction, which can be achieved only through a full next-to-leading
  order (NLO) calculation.  
  In this work we describe the NLO computation of the 
  subprocess $gg\to t \bar t g$.}
\end{center}

\begin{abstract}
The reaction $pp/p\bar p \to \ttbarjet+X$ is an important background
process for Higgs boson searches in the mass range below 200~GeV. Apart
from that it is also an ideal laboratory for precision measurements in
the top quark sector. Both applications require a solid theoretical
prediction, which can be achieved only through a full next-to-leading
order (NLO) calculation.  
In this work we describe the NLO computation of the 
subprocess $gg\to t \bar t g$.
\end{abstract}
\maketitle

\section{Introduction}
The main objective of the Large Hadron Collider (LHC) at CERN is
the discovery of the Higgs boson and the measurement of its mass and couplings.
To achieve this important goal a solid knowledge of the production
mechanisms and the corresponding backgrounds is mandatory. In the
Standard Model a light Higgs boson is currently favoured by the 
available data. 
Using the recently
updated top mass of $m_t = 178.0 \pm 4.3$~GeV \cite{Azzi:2004rc} 
the electroweak fits yield an upper bound of 251~GeV (at~95\%~C.L.) 
and a central
value of $\mhiggs = 117$~GeV \cite{SRoth-Moriond}.
To achieve a high signal significance in the Higgs searches 
in general, different production and decay mechanisms are combined. 
In the range up to 200~GeV the so-called weak boson
fusion (WBF) process with the subsequent decay of the Higgs into a W-boson
pair  plays a dominant r\^ole. The most important background for the
WBF process comes from the \ttbarjet\ process \cite{Alves:2003vp}. 
A very precise knowledge of this process is thus 
mandatory for the discovery of the Higgs boson. It is obvious that for precise
measurements of the couplings a precise background
determination is equally important. For example, 
it has been shown in \Ref{Rainwater:2002hm} 
that even if one assumes only a 10\% uncertainty of the 
\ttbarjet\  cross section it is still the dominant theoretical
uncertainty in the measurement of 
$\sigma_{\rm H}=\sigma_{\rm WBF} \times B(H\to WW)$.
As also pointed out in \Ref{Rainwater:2002hm} 
this accuracy might be achievable only 
through a full next-to-leading order (NLO) calculation. In a recent
analysis \cite{Cavalli:2002vs,Kauer:2004fg}, the possibility 
to extract the background from
extrapolation of experimental data has been studied. In this analysis
it was found that a background determination with 5--10\% accuracy
might be possible. This is a very promising result. On the other hand
--- given the significance of the precise background determination ---
we believe that a cross-check with a full NLO QCD prediction is
important. At the very end --- having a good understanding of both
results --- both methods/results could and should be used as complementary.

In fact the \ttbarjet\ reaction is not only important as background
for Higgs searches, it is also an important signal process on its own. 
It is well known that top quark physics allows a test of the Standard Model at 
high scale. In particular one can  search for possible extensions
of the Standard Model at the scale of the top quark mass.
As far as top quark production at hadron colliders is concerned, 
the state of the art is as follows. The
differential cross section for top quark pair production is known to 
next-to-leading order accuracy in QCD 
\cite{Nason:1988xz,Nason:1989zy,Beenakker:1989bq,Beenakker:1991ma}.
In addition the resummation of logarithmically enhanced contributions has
been studied in detail in 
\Refs{Laenen:1992af,Laenen:1994xr,Kidonakis:1995wz,Berger:1996ad,Catani:1996yz,Kidonakis:1997gm,Berger:1998gz,Bonciani:1998vc}.
Recently also the spin correlations between top quark and antitop quark
were calculated at NLO in QCD \cite{Bernreuther:2001rq,Bernreuther:2004jv}.
Since single top quark production provides an excellent 
opportunity to test the charged-current weak interaction of the top
quark, it has also attracted a lot of interest in the past. In
particular NLO corrections were studied in 
\Refs{Bordes:1995ki,Smith:1996ij,Stelzer:1997ns,Harris:2002md}. In 
\Ref{Harris:2002md} the NLO corrections for the fully
differential cross section are given, keeping also the spin information 
of the top quark.

In that context the natural next step is the calculation of the NLO 
corrections for \ttbarjet\ production.   As far as top quark physics
is concerned, interesting observables to study are those that vanish 
if there is no additional jet. Such observables allow for a direct test 
of the dynamics in the top quark sector. For example the 
asymmetry \cite{Rainwater}
\begin{equation}
  A(y) = { N(\ell,{\rm forward}) -N(\ell,{\rm backward})\over 
    N(\ell,{\rm forward}) +N(\ell,{\rm backward})} 
\end{equation}
is such an observable. Here $ N(\ell,{\rm forward/backward})$ 
denotes the number of forward/backward-going leptons as a function 
of the rapidity $y$ of the additional
jet. The measurement of this asymmetry allows for a direct test of the
production and decay mechanisms.
Using similar observables one can search for example
for anomalous top--gluon couplings. A more precise understanding of
the cross sections for $pp/p\bar p \to t \bar t + \mbox{jets} +X$ 
is also important for measurements of the top quark mass.

\section{Outline of the calculation}
In this section we briefly summarize the calculation of the NLO corrections
for the subprocess $gg\to t\bar t g$. 
In view of the number of external legs
and the top mass as additional parameter it is obvious that even 
partial results are in general quite lengthy. 
In the following we will restrict our attention only to those parts
of the calculation
where special care is needed to construct a numerically stable program. 

\subsection{Virtual corrections}
The calculation of the  virtual corrections proceeds via the following
steps:
\begin{enumerate}
\item Generation of the Feynman diagrams using for example 
  Feynarts \cite{Kublbeck:1990xc} or QGRAF \cite{Nogueira:1993ex}.
\item Reduction of the tensor integrals to scalar one-loop integrals.
\item Reduction of the amplitudes to standard matrix elements.
\item Numerical phase-space integration of the squared 
  matrix elements, including appropriate phase-space cuts.
\end{enumerate}
Technically the most complicated part is the evaluation of the
pentagon-diagrams.
\begin{figure}[htbp]
  \begin{center}
 \includegraphics[width=3.cm]{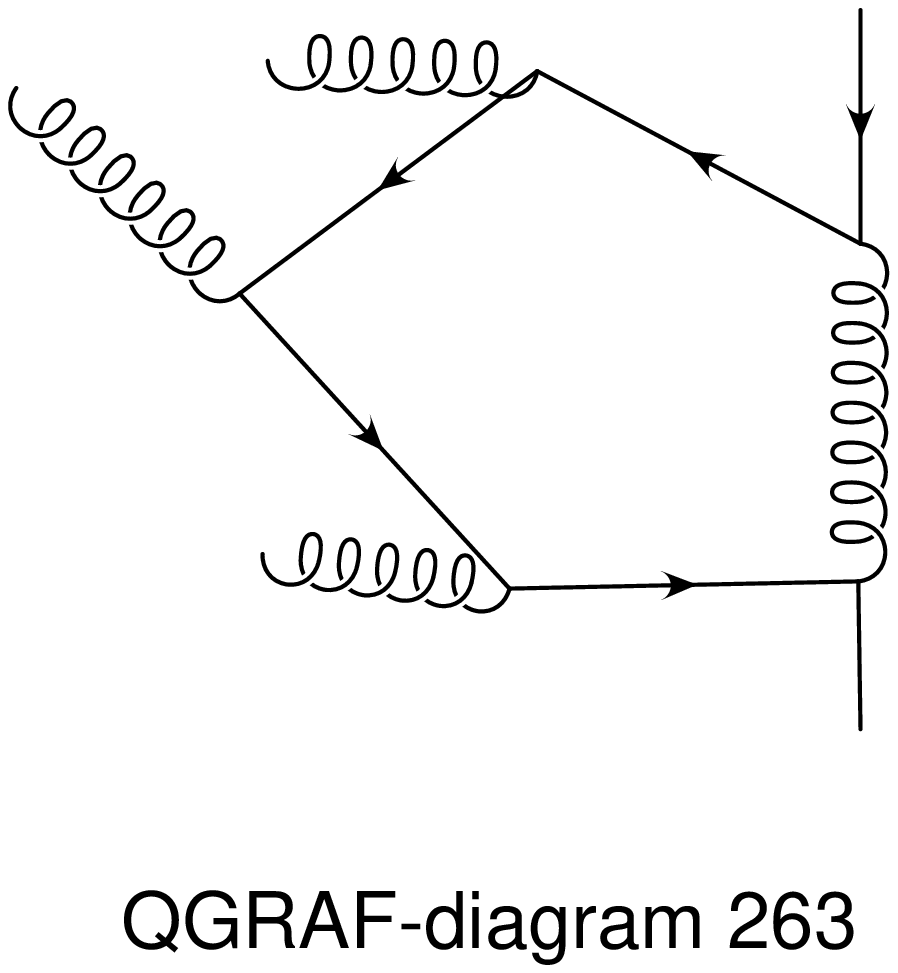}\hfill
 \includegraphics[width=3.cm]{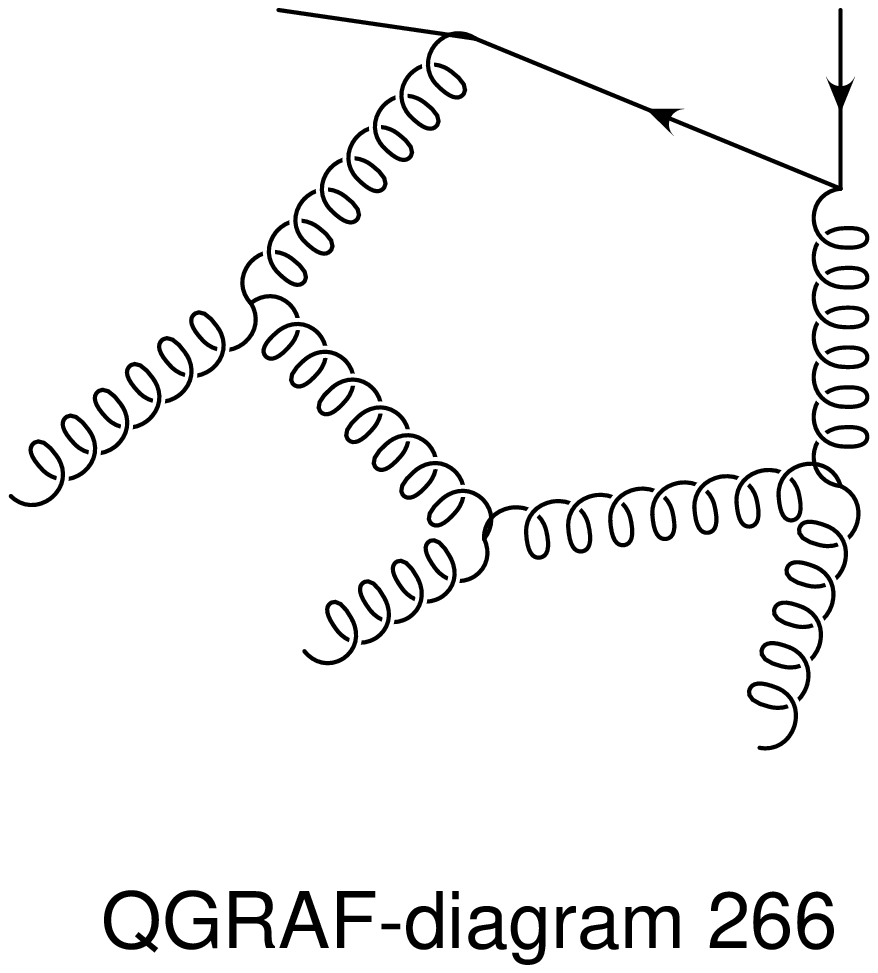}
    \caption{Sample Feynman diagrams contributing to the virtual corrections. 
      \label{fig:Virtual}}
  \end{center}
\end{figure}
Two sample diagrams are shown in \Fig{fig:Virtual}. Let us first
address the evaluation of the scalar 5-point integrals. To calculate
these, we use two different methods. One
calculation is based on the method given in
\Refs{Beenakker:2001rj,Beenakker:2002nc}. 
The basic idea of this method is that finite
5-point integrals can be expressed in terms of 4-point integrals (see
for example \cite{Melrose:1965kb,vanNeerven:1984vr,Bern:1994kr}). To
apply this observation also to soft- and mass-singular integrals they
are rewritten according to \Refs{Beenakker:2001rj,Beenakker:2002nc} 
in the following way: 
\begin{equation}
   E^d = E^{d}_{\rm sing.} + \left[ E^{({\rm mass},d=4)}
     -E^{({\rm mass},d=4)}_{\rm sing.}\right].
\end{equation}
Here $E^d$ denotes the original 5-point integral in $d$ dimensions
while  $E^{({\rm mass},d=4)}$ is obtained from the original integral
by dressing the massless propagators with a small mass $\lambda$.  
The subtraction term $E^{({\rm mass},d=4)}_{\rm sing.}$, which
has the same singular structure as the 5-point integral 
$E^{({\rm mass},d=4)}$ in the limit $\lambda \to 0$, is
obtained by studying the soft and collinear behaviour of 
$E^{({\rm mass},d=4)}$
and can be expressed in terms of 3-point functions \cite{Dittmaier:2003bc}.  
Rewriting now the finite integral $E^{({\rm mass},d=4)}$ in terms
of 4-point integrals we thus succeeded
in expressing the original 5-point integral in terms of 
3- and 4-point functions. 
A more detailed discussion can be found in \Ref{Beenakker:2002nc}.
The second method we used to calculate the five-point integrals is
based on the fact that, even for divergent integrals, it is possible to
obtain a representation as linear combination of 4-point
integrals (see for example \Ref{Bern:1994kr}). 
Expressing the 4-point function for $d=4-2\epsilon$ in terms of the 
finite 4-point function
in 6 dimensions plus a combination of 3-point integrals allows us also
to
shift all the divergences to the 3-point integrals.
Defining the 5-point functions through
\begin{eqnarray}
  &&E^d(p_0,p_1,p_2,p_3,p_4,m_0,m_1,m_2,m_3,m_4)\nn\\
  &=&
  {1\over i\*\pi^2}
  \int d^d\ell \prod_{j=0}^{4}{1\over (\ell+p_j)^2-m_j^2+i\epsilon},
\end{eqnarray}
we obtain for example 
\begin{eqnarray}
  &&\hspace{-1cm}
  \left.E_0(0,p_1,p_1-p_3,p_4-p_2,-p_2,m_t,m_t,0,0,m_t)\right|_{\rm
    sing.} \nn \\
  &=& 
  P(t_{13})P(s_{45})\*C_0(p_1-p_3,p_4-p_2,p_1,0,0,m_t) \nn \\
  &+&
  P(t_{24})P(s_{35})\*C_0(p_4-p_2,-p_2,p_1-p_3,0,m_t,0)\nn \\
  &-& 
  (t_{13}-t_{24})^2 P(t_{13})P(t_{24})P(s_{35})P(s_{45})\nn\\
  &&\times C_0(0,p_1-p_3,p_4-p_2,m_t,0,0),
\end{eqnarray}
with $P(x) = 1/(x-m_t^2)$ and $s_{ij} = (p_i+p_j)^2$, 
$t_{ij} = (p_i-p_j)^2$. The parton momenta are assigned according to 
$g(p_1)g(p_2) \to t(p_3)\bar t(p_4) g(p_5)$.
For the cases at hand it is possible to solve all the required
box-integrals in 6 dimensions. 
We checked that the two methods yield the same results for the 5-point 
integrals $E^d$.

Having solved the scalar integrals, the next step is the reduction of
the 5-point tensor integrals to scalar one-loop integrals. 
In principle one could attack
this problem using the standard Passarino--Veltman approach 
\cite{Passarino:1979jh}. 
This method leads to spurious singularities in
individual terms at the
phase-space boundary due to vanishing Gram determinants in the
denominator.  
These spurious singularities create
numerical instabilities when doing the phase-space integration. 
Note that the spurious singularities cancel if one combines the
individual terms analytically before doing the numerical integration.
One solution of  this problem is a time-consuming
extrapolation technique, as was used for example in
\Ref{Beenakker:2002nc}. As an alternative to the extrapolation
technique a different reduction procedure \cite{Denner:2002ii} was also
used  in \Ref{Beenakker:2002nc}.
In this work we
follow the method developed in \Ref{Denner:2002ii}.
Essentially the same technique to reduce scalar 5-point integrals to scalar
4-point integrals is also applied to the tensor integrals. In this way the 
5-point tensor integrals are directly reduced to 4-point tensor
integrals. The explicit calculation shows that in this way the
spurious singularities in individual terms, due to vanishing Gram
determinants depending on 4 external momenta, are avoided \cite{Denner:2002ii}.

Let us just mention at the end that there are also other methods to solve 
the scalar 5-point integrals and perform the reduction of the tensor 
integrals. For example one could also use the methods developed in 
\Refs{Giele:2004iy,Binoth:1999sp,Duplancic:2003tv}. 
(For \Ref{Giele:2004iy} see also Walter Giele's talk in these proceedings.) 

\subsection{Real corrections}
The calculation of the required matrix elements is straightforward. 
\begin{figure}[htbp]
  \begin{center}
    \includegraphics[width=3.cm]{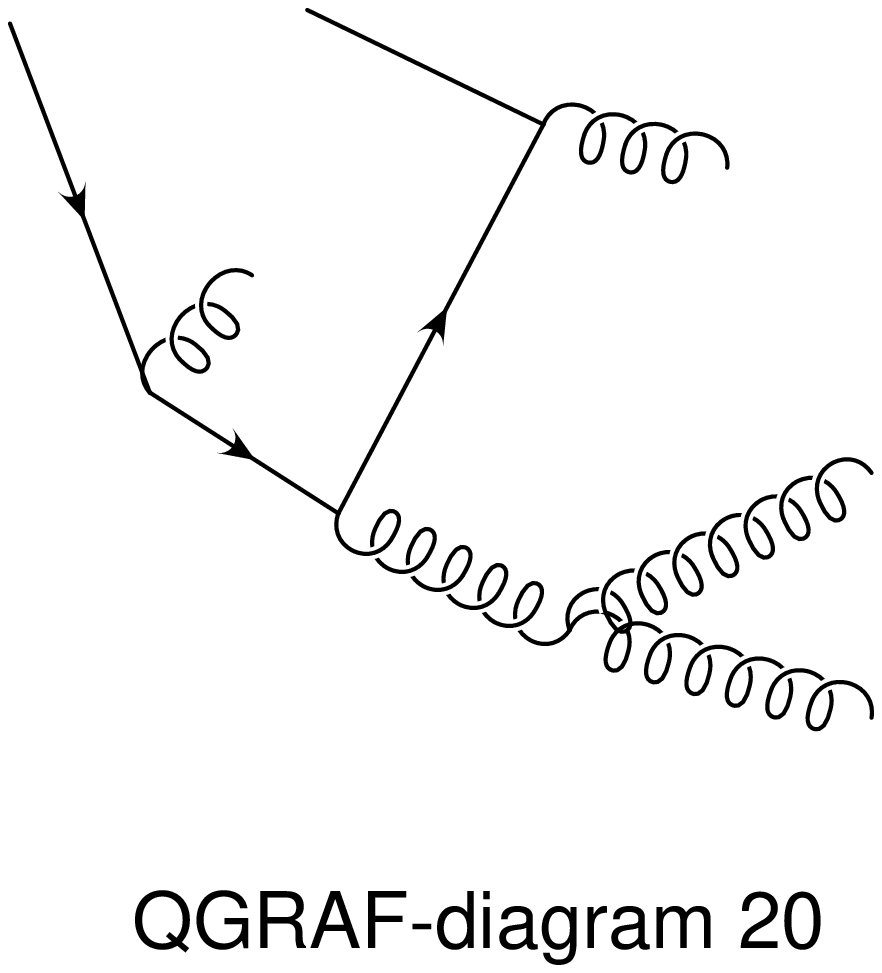}
    \caption{Sample Feynman diagram contributing to the real corrections. 
      \label{fig:Real}}
  \end{center}
\end{figure}
A sample diagram for the reaction $gg \to t \bar t gg$ is shown in 
\Fig{fig:Real}. We used two different methods to obtain the required
colour-ordered helicity amplitudes:
\begin{enumerate}
\item A Feynman-diagram-based approach where we evaluate all the
  diagrams contributing to one specific colour-ordered subamplitude.
\item Using the recurrence relations à la Berends and Giele 
\cite{Berends:1988me}.
\end{enumerate}
We find complete agreement in the results of the two methods. 
Furthermore we also checked 
that our results agree with the ones obtained using 
Madgraph \cite{Stelzer:1994ta}.
To extract the singularities from collinear or soft partons we use the
dipole subtraction method \cite{Catani:1997vz,Phaf:2001gc,Catani:2002hc}.
The idea of the subtraction method is to add and subtract a term which,
on the one hand, cancels pointwise the singularities of the
matrix elements in the singular regions of the phase-space and is, on the
other hand, easy enough to be integrated analytically. Schematically
the NLO contribution is then obtained from the following formula:
\begin{eqnarray}
  \label{eq:SubtractionMethod}
 \sigma_{\rm NLO}
  &=& 
  \underbrace{\int_{m+1}\left[\sigma_{\rm real}-\sigma_{\rm sub}\right]}_{\rm
    finite}
  +
  \underbrace{\int_{m}\left[\sigma_{\rm virt.}+\int_1 \bar\sigma^1_{\rm
  sub}\right]}_{\rm finite}\nn\\
  &+&
  \underbrace{\int dx\int_{m}\left[\sigma_{\rm fact.}(x)
    +\bar\sigma_{\rm sub}(x)\right]}_{\rm finite}.
\end{eqnarray}
Here $\sigma_{\rm fact.}(x)$  denotes the contribution from the 
factorization of initial-state singularities due to the presence of 
coloured partons in the initial
state. The contributions $\bar\sigma^1_{\rm
  sub},\bar\sigma_{\rm sub}$ are obtained from $\sigma_{\rm sub}$
by integrating out the `unresolved' parton. The result is split
into the two terms $\bar\sigma^1_{\rm
  sub},\bar\sigma_{\rm sub}$ to render the last two integrals
individually finite. A remarkable feature of the subtraction method is
that the analytic integration of the subtraction has to be done only
once and that in the whole procedure no approximation is made.
This is made possible by the universality of soft and collinear
factorization in QCD. The explicit expressions for $\sigma_{\rm sub},
\sigma^1_{\rm sub},$ and $\bar\sigma_{\rm sub}$ can be obtained from
the colour-ordered subamplitudes using the formulae  given in 
\Ref{Catani:2002hc}. In particular $\sigma_{\rm sub}$ is obtained
from a sum over individual {\it dipole contributions}. In the case at hand we
have to include the contribution from 36 individual dipoles. We do not
consider the splitting $g\to t\bar t$ because the divergence is
regulated by the quark masses. (For light quarks one could consider the
corresponding dipoles to render the integration numerically more stable.)
We have checked that the combination of the 36 dipoles indeed
reproduces all the singular limits arising from single unresolved 
configurations.

\section{Status and results} 
The current status of the project is as follows. Most of the separate
contributions are implemented in the form of  computer
programs allowing the numerical evaluation of the cross sections.
\begin{figure}[htbp]
  \begin{center}
 \includegraphics[width=7.5cm]{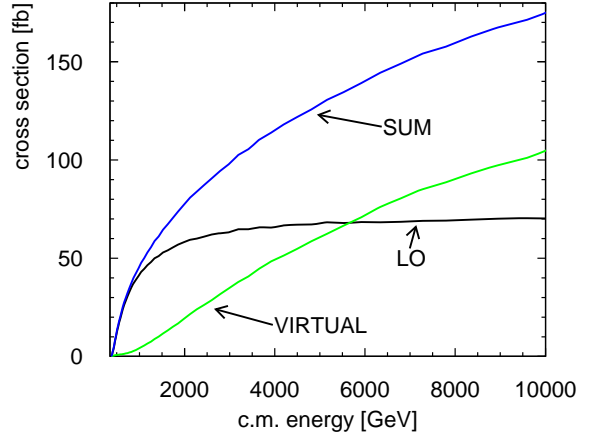}
    \caption{ Result for the virtual corrections for the 
      subprocess $gg\to t\bar t g$
       as defined by the
      second term in \Eq{eq:SubtractionMethod}\  ($k_\perp>$~20~GeV). 
      \label{fig:ResultVirt}}
  \end{center}
\end{figure}
In \Fig{fig:ResultVirt} we show as an example, the result at the parton level
for the virtual
corrections (defined as the second term in \Eq{eq:SubtractionMethod})
for different centre-of-mass energies. 
(Given that the separation shown in \Eq{eq:SubtractionMethod} 
involves some freedom, this individual contribution does not have a
direct physical interpretation
unless the remaining contributions are added --- we just show it for
illustrative purposes.) As can be seen from \Fig{fig:ResultVirt} the
method we used for the treatment of the tensor integrals gives indeed
numerically stable results.
Furthermore we note that the inclusion of $d\bar \sigma^1_{\rm sub}$
together with the renormalization of the coupling and the quark mass
renders the second term in \Eq{eq:SubtractionMethod} 
finite, as it must be. This is an important cross-check.
As mentioned earlier we also checked  that the integrand for the 
first contribution in \Eq{eq:SubtractionMethod} is also finite for all 
single unresolved phase-space configurations.
Given the complexity of the project we think it is very important to 
have independent cross-checks for every individual contribution. 
While most of the calculation is already cross-checked, we  still
work to finish also the remaining checks. Complete results will
be presented elsewhere.

\section{Conclusions} 
In this work we discuss the NLO calculation for the partonic
reaction $gg\to t \bar t g$. Up to remaining cross-checks, the
calculation is almost finished. In particular we have shown that the 
virtual corrections are stable using the reduction procedure discussed
in \Ref{Denner:2002ii}.

{\bf Acknowledgements:} We would like to thank the organizers for the
pleasant atmosphere at Loops and Legs 2004.

\newcommand{\zp}{Z. Phys. }\def\as{\alpha_s }\newcommand{\prd}{Phys. Rev.
  }\newcommand{\pr}{Phys. Rev. }\newcommand{\prl}{Phys. Rev. Lett.
  }\newcommand{\npb}{Nucl. Phys. }\newcommand{\psnp}{Nucl. Phys. B (Proc.
  Suppl.) }\newcommand{\pl}{Phys. Lett. }\newcommand{\ap}{Ann. Phys.
  }\newcommand{\cmp}{Commun. Math. Phys. }\newcommand{\prep}{Phys. Rep.
  }\newcommand{\jmp}{J. Math. Phys. }\newcommand{\rmp}{Rev. Mod. Phys. }

\end{document}